\documentclass[runningheads]{llncs}
\usepackage{graphicx}
\usepackage{subfigure}
\usepackage{booktabs}
\usepackage{pifont}
\usepackage[misc]{ifsym}
\usepackage[symbol]{footmisc}

\usepackage[colorlinks,
            linkcolor=black,
            anchorcolor=black,
            citecolor=black]{hyperref}
\begin{document}
\begin{sloppypar}
\title{PCDF: A Parallel-Computing Distributed Framework for Sponsored
Search Advertising Serving}
%
%
\author{Han Xu \footnotemark[1]\orcidID{0009-0006-3264-1226} \and
Hao Qi \footnotemark[1] \and
Yaokun Wang\and
Pei Wang\footnotemark[2]\orcidID{0000-0001-9910-4114}\Letter  \and
Guowei Zhang\footnotemark[3]\orcidID{0000-0002-3288-9946}\and
Congcong Liu\orcidID{0000-0002-1749-1075}  \and
Junsheng Jin\and
Xiwei Zhao\and
Zhangang Lin\and
Jinghe Hu\and
Jingping Shao
}

\renewcommand{\thefootnote}{\fnsymbol{footnote}} 
\footnotetext[1]{Equal contribution.} 
\footnotetext[2]{Corresponding author.} 
\footnotetext[3]{The author made a lot of contributions to this work.}

%
\institute{JD.com, Beijing, China \\ 
\email{xhbj66@gmail.com,\{qihao1,wangkunyao\}@jd.com,1640715678@qq.com,wangpei102595@gmail.com,
cliubh@connect.ust.hk,\{
jinjunsheng1,zhaoxiwei,linzhangang,hujinghe,shaojingping\}@jd.com}}

\authorrunning{Han Xu, Hao Qi, Kunyao Wang, Pei Wang, Guowei Zhang, et al.}
\titlerunning{Sponsored Search Ads Distributed Framework}
\toctitle{Sponsored Search Ads Distributed Framework}
\tocauthor{Han~Xu}
\tocauthor{Hao~Qi}
\tocauthor{Yaokun~Wang}
\tocauthor{Pei~Wang}
\tocauthor{Guowei~Zhang}
\tocauthor{Congcong~Liu}
\tocauthor{Junsheng~Jin}
\tocauthor{Xiwei~Zhao}
\tocauthor{Zhangang~Lin}
\tocauthor{Jinghe~Hu}
\tocauthor{Jingping~Shao}
\maketitle              
%
\begin{abstract}
Traditional online advertising systems for sponsored search follow a cascade paradigm with retrieval, pre-ranking,ranking, respectively. Constrained by strict requirements on online inference efficiency, it tend to be difficult to deploy useful but computationally intensive modules in the ranking stage. Moreover, ranking models currently used in the industry assume the user click only relies on the advertisements itself, which results in the ranking stage overlooking the impact of organic search results on the predicted advertisements (ads). In this work, we propose a novel framework PCDF(Parallel-Computing Distributed Framework), allowing to split the computation cost into three parts and to deploy them in the pre-module in parallel with the retrieval stage, the middle-module for ranking ads, and the post-module for re-ranking ads with external items. Our PCDF effectively reduces the overall inference latency compared with the classic framework. The whole module is end-to-end offline training and adapt for the online learning paradigm. To our knowledge, we are the first to propose an end-to-end solution for online training and deployment on complex CTR models from the system framework side.


\end{abstract}
\section{Introduction}\label{introduction}
CTR prediction is the core task of advertising systems, predicting the probability of the users' click events on a certain item. A typical paradigm in online advertising systems is to retrieve a subset of advertisements relevant to the users from a large corpus by a candidate generation network\cite{ref_nework}, and then rank these candidates through a ranking network\cite{ref_nework,ref_youtube}, leaving only a few items to present to the user.

Driven by the advancement of deep learning, large-scale deep neural networks are usually employed as ranking models in recommendation systems to achieve good system performance. 
However, complex models are difficult to deploy under extremely low system latency constraints in real-time recommendation systems. 
Many existing works focus on improving the effectiveness and efficiency of recommender systems\cite{ref_youtube,ref_Clipper,ref_large,ref_airbnb,ref_RecNMP,ref_pipedream,liu2022position,zhu2023confidence,liu2023always,liu2022concept,zhu2021dynamic,liu2022rethinking}. 
Recent works \cite{ref_indexing,ref_cache-center,ref_UIC,ref_user_retrieval}reduce computational latency and improve system efficiency by using two-stage modeling methods.
Although these methods reduce the computation cost in the ranking stage, maintaining data consistency between the two-stage brings the challenge to the online serving system, and the ranking model doesn't fully exploit the rich information contained in the features generated in the one-stage model. Furthermore, these methods are not suitable for the online learning paradigm, which further impairs the accuracy of prediction.
The above methods concentrate on pre-reducing online inference time through model design in the ranking stage but ignore the rationality of the overall framework.

From the system deployment side, many existing recommendation systems decompose models and data to benefit from data parallelism and model parallelism\cite{ref_task_data,ref_model_para,ref_model_para1}. However, the acceleration revenue of the above two strategies reaches a bottleneck since model computing complexity keeps growing.  Pipeline parallelism\cite{ref_pipedream} has been a handful solution to simplify the design of algorithms and facilitate deployment, but many recommendation frameworks applied pipeline parallelism suffer from the unbalanced load.

Given the above limitations, in this paper, we rethink the challenges of deploying complex models in e-commerce search platform, from a system design perspective and take user long-term behavior modeling and organic search information modeling as the case study. 
Unlike other methods that design training-inference inconsistent models to reduce online computational complexity, we adopt end-to-end training and maintain training-inference consistency to achieve good performance.
More specifically, We split the precision ranking model into three modeling stages: pre-modeling, middle-modeling, and post-modeling, and perform parallel computing for pre-modeling and post-modeling with other modules in the recommendation system.
Benefiting from the parallel framework, the latency of the whole ranking model can be reduced even if adding a complex target-independent module. The main contributions of this work are summarized as follows:
\begin{itemize}
    \item We propose a novel Parallel-Computing Distribution Framework(PCDF) from the perspective of system framework design. 
    To the best of our knowledge, PCDF is the first systematic solution for deploying computationally expensive target-independent modules in online advertising serving. 
    \item A new pipeline parallelism recommendation inference strategy is proposed, which split the deep rank stage into three modeling stages: pre-modeling, in-modeling, and post-modeling; the above three-stage deep rank process is carried out in a pipeline parallelism way by one single deep rank model. Computing power cost in the ranking stage is explicitly reduced by applying  PCDF. This further brings space for the advertising systems to apply more complex deep models to reach better performance.
    \item We introduce a hands-on practice of the PCDF framework on both offline training and online deployment for CTR prediction on a real-world advertising platform. 
    \item The modeling for Long-term user historical behavior and  externalities is deployed as a task and conduct comprehensive offline and online experiments to validate our solution's rationality. We achieve a 5\% improvement on CTR and 5.1\% improvement on RPM  in the online A/B test. And there is almost no increase in inference time in  the whole ranking system.
   
\end{itemize}

\section{Related Work}
\subsection{ Latency Optimization}
There are generally several methods for optimizing the performance of online inference.  On the model side, two common methods are model pruning and model quantization with low-precision inference. Model pruning reduces model size by removing unnecessary weights or neurons, which speeds up inference. Model quantization converts floating-point model parameters to integers or uses lower bit-width data representation to accelerate inference.\cite{polino2018model,zhou2018adaptive,lin2020dynamic,jiang2022model} However, these methods may lead to a reduction in model parameters, which can decrease the model’s accuracy and performance. Achieving optimal results requires significant experimentation and computational resources to adjust parameters. On the framework side, parallel computing distributes the model or data across different nodes for parallel modeling training, however, the parallel computing approach does not conform to the paradigm of online learning and is rarely used in model inference.

Once the traditional three stages have been carried out in e-commerce search platform, a post-processing module is typically deployed. This module is utilized for the reordering of strategies between items, as well as for handling new strategies resulting from different business practices. Due to system latency limitations, it is generally difficult to deploy complex models in the post-processing stage. In order to cut down on the latency it takes to run the post-processing module, optimization methods such as edge computing are used to eliminate latency from the edge to the server\cite{gong2020edgerec,pustokhina2020effective}. Additionally, performance-enhancing algorithmic strategies are applied to decrease computation cost\cite{pei2019personalized,feng2021revisit}.


To address the aforementioned issues, we propose a pipelined parallel framework that divides the model inference process into three stages: pre-modeling, middle-modeling, and post-modeling. Through pipeline parallelism and parallel computing, we successfully deployed complex models without increasing model latency. This approach improved model inference accuracy and achieved good online performance benefits.

\section{Methodology}
In this section, we explain the motivation behind the PCDF (Parallel Computing Distributed Framework) framework and introduce its design considerations. We then provide a detailed overview of the architecture of the deployment framework, summarized the online serving framework of the e-commerce sponsored search system, and introduce our proposed PCDF. Next, we briefly explain how the models, including behavior modeling and external information modeling, are applied in PCDF  to demonstrate that PCDF provides scalable, efficient, and easy-to-use deployment solutions for machine learning models.

\subsection{Preliminaries}
\subsubsection{CTR Task.} \label{ctr_task}
The CTR prediction is to predict the probability that a user clicks an item. After retrieving hundreds of candidate items at the retrieval stage, a ranking module is employed to make a prediction between user $u$ and every item $x$ from candidate items $X$. In general, a CTR prediction model involves four kinds of features:
\begin{equation}
    CTR = f(L_u^{1:T_l},S_u^{1:T_s}, x_u, x_t, x_c)
\end{equation}
where $L_u^{1:T_l}$ and $S_u^{1:T_s}$ are long and short historical behaviors respectively, $x_u$, $x_t$, $x_c$ corresponds to user profile, target item profile and contexts. 
As shown in Fig\ref{fig:subfeature1}, First, the user request $R_u$ from upstreams is handled by a retrieval Module for hundreds of items $I_R$ this user $u$ might be interested in. Followed by predictions on all items made by Pre Rank Module and Deep Rank Module. Each module has a Feature extract process, to process all Module required features ($L_u^{1:T_l},S_u^{1:T_s}, x_u, x_t, x_c$). Since  Rank Modules require the candidate items responded from Retrieval Module, the whole process is serialized.
\begin{figure}[htbp]
\centering
\includegraphics[width=1\textwidth]{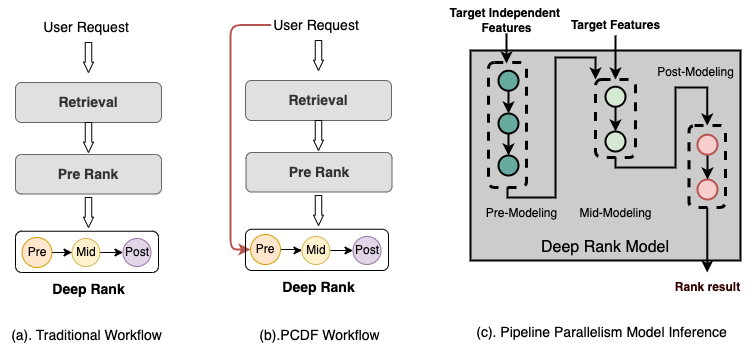}
\caption{Illustration of traditional recommendation workflow and PCDF workflow, traditional recommendation workflow illustrated in fig.{\bfseries (a)} includes three stages: retrieval, pre ranking, and deep ranking. The deep ranking stage includes three aspects of work: target-independent pre preprocessing, model estimation, and post-processing.PCDF workflow illustrated in fig.{\bfseries(b)} adds a pre-computing stage to move the target-independent processing of the fine ranking stage forward in parallel with the recall and pre ranking, which reduces the burden of the deep ranking stage.
in fig{\bfseries(c)}, the deep ranking model is split into three stages: pre-modeling, mid-modeling, and post-modeling, which can be performed in parallel to support target-independent and target-dependent feature computation, as well as post-processing.}
\label{fig:workflow}
\end{figure}
\subsubsection{E-Commerce Search Platform.}
A practical e-commerce search platform that recommends multiple products to online users, usually through two separate systems, the Organic Search System (OS) and the Sponsored Search Advertisement System (AS). To present a mix of search and advertising listings, the organic results The list is first generated by the OS, and then the AS assigns the correct ad on the correct ad slot\cite{ref_os}. In this paper, we mainly focus on the search advertising system, which follows a three-stage design with recall, rough sorting, and sorting. Specifically, a large number of items and Ads that are relevant to the user are first retrieved from the query, which is then sent to a predictive model that estimates various ad quality metrics, such as click-through rate. These candidate ads are then ranked by the metrics generated and advertiser bids. The winning bid after these auction advertisements will ultimately be presented to consumers along with organic search results\cite{ref_youtube,ref_indexing,ref_sim}.


\subsection{Design Considerations And Motivation}\label{sec:consider}
{\bfseries 1) Strict Online Serving Latency.} The major target for designing a real-time recommendation system is to reach peak recommendation accuracy under strict constraints of latency.  Generally speaking, as prediction accuracy increases with model complexity, the  trade-off between prediction accuracy and system latency has to been made, due to the limitation of latency.  In our scenario with heavy throughput, there are lots of optimization work, including multi-thread, data parallelism, and tensorflow op parallelism to keep system latency under 60ms which is the system latency of our recommendation system.

{\bfseries 2) Deployment Costs.} 
Given that GPU has a good acceleration effect on many models, currently
mainstream recommendation system will use GPU for inference acceleration while the cost of GPU resources is higher than that of CPU. Many recommendation system architectures deploy CPU resources and GPU resources together, however, it is difficult to accurately adjust the allocation ratio of GPU computing resources, bandwidth, and CPU computing resources; with the iterative changes of the model, an unbalanced load of CPU and GPU leads to higher deployment costs leading to high costs for large-scale commercial recommendation system.

We formulate Large-scale Distributed Real-Time Prediction (RTP) system in Figure \ref{fig:workflow}(a), which consists of three core components: Retrieval module, Pre Rank Module, and Deep Rank Module. The computation latency is generally unsatisfying if complex model 
is performed on thousands of advertisements in the retrieval stage\cite{ref_youtube}. Thus complex model 
is usually deployed and performed in the deep rank stage after ads are filtered in the retrieval stage. As introduced in section\ref{ctr_task}, the traditional three-stage recommendation process is serialized. However, not all features are only accessible after the retrieval Module like target-independent features ($L_u^{1:T_l}$, $S_u^{1:T_s}$, $x_u$ and $x_c$) since we can know the user and context information from the initial user request. We propose a stage-level parallel computing idea shown in Figure \ref{fig:workflow}(b) to reduce online reasoning time. Show in Figure \ref{fig:workflow}(c), pipeline parallelism is used to support the splitting of the deep rank stage process.

\begin{figure}[htbp]
\includegraphics[width=\textwidth]{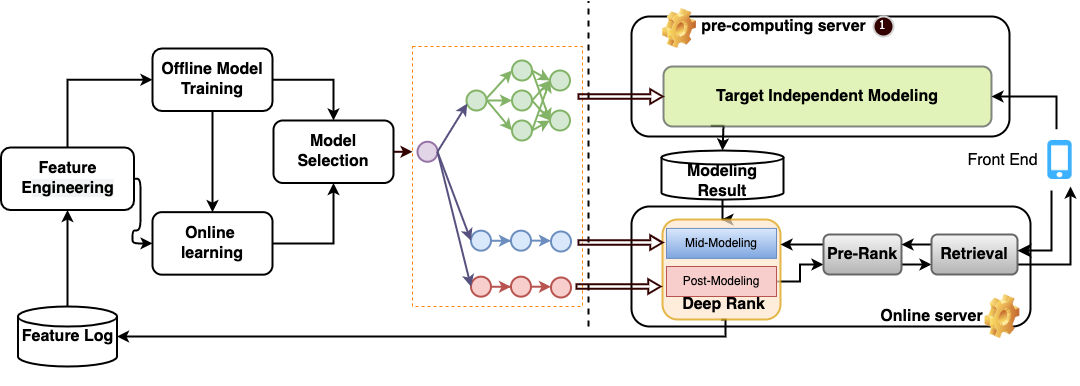}
\caption{Illustration of online advertising system. Typically it consists of three important components: real-time serving module, offline model training, and model deploy module. Left is the feature log module and Model training module, after model training, the new model will be split and deployed in a real-time server. The same model will perform different calculations at different stages to achieve a staged pipeline model inference process.} \label{fig:architecture}
\end{figure}

\subsection{PCDF Framework} \label{subsction:framework}
In this section, we will introduce our newly designed ranking system PCDF in detail. The core idea behind PCDF is to take into consideration of both model design and system design. The high-level architecture of PCDF is illustrated in Figure \ref{fig:architecture}. PCDF consists of two components: the training component(shown in the left, which contains model online learning, Feature Engineering, and model detachment module) uses Tensorflow to train the model serving for online server, and data used for training model comes from user and target features.


\subsubsection{Deep Ranking Model in PCDF.}\label{subsction:model}
As shown in Figure \ref{fig:architecture}, model designed for online serving is divided into three sub-models (pre-model, mid-model, post-model) and deployed on an online server, where the online serving component recommends items to users based on their preferences in real-time. Specifically, when a front-end user requests recommended targets, it obtains task-independent features and sends the processed features to the pre-processing model. After the pre-processing stage is completed, the captured information is sent back to the deep ranking module. These processes run in parallel with the retrieval module’s call. Once the deep ranking module receives feedback from the pre-ranking and pre-processing modules, it reprocesses the features to obtain item features and sends all the results to the ranking module in the deep ranking model. As the main ranking stage, the mid-model and post-model rank all candidate items based on item features, user features, context features, and externalities in e-commerce sponsored search.

Shown in figure \ref{fig:workflow}(c),we design a pipeline parallelism strategy that allows the same model to provide modeling services for different stage modeling processes in the deep ranking stage. Firstly, pre-modeling is used to model target-independent features in the first stage. Then, the outputs of pre-modeling along with target features are sent to the mid-modeling process to obtain prediction scores. Finally, a post-modeling is applied to process the mid-modeling results and externality. Next, we will introduce how PCDF is deployed in the online learning and serving manner.In industry, online learning refers to updating the model parameters in real-time through online requests during the online service stage, in order to adapt to the dynamic distribution of data.

{\bfseries Training. } As shown on the left side of Figure \ref{fig:architecture}, 
after the target-independent modeling is finished, feature engineering is performed on the feature log with hadoop\cite{ref_hadoop}, including fusing the outputs of the pre-computing server with other features related to candidate items.
In the way that all modules in the ranking model are jointly learning to optimize the deep ranking task, the parameters of the part for the target-independent model are synchronously updated with other parts through training in an end-to-end way. 

The model training process adopts the online learning paradigm to update the model. With the continuous generation of data streams, the model can dynamically update its parameters to adapt to the new data distribution as new data arrives. During this process, the model’s training and inference are performed alternately, and the model can continuously learn and improve from new data. The model selection module will prioritize the selection of different model and push it to the real-time recommendation system to start providing services. 
Joint learning enables  the information of target-independent features fully utilized, compared to the two-stage models\cite{ref_sim,ref_user_retrieval}.
Besides, the end-to-end model updated all parameters synchronously, which keeps consistency during online learning.

{\bfseries Pipeline Parallelism Servering.}  Deep Ranking stage on the right of Figure \ref{fig:architecture} is divided into three stages: pre-modeling, mid-modeling, and post-modeling.
{\bfseries 1) Pre-Modeling.} target independent pre-modeling process is deployed in pre-computing module\ding{182} and triggered simultaneously with retrieval process by front end user request. The results of pre-modeling are cached by redis\cite{ref_redis}. {\bfseries 2) Mid-Modeling.} After the processes of the retrieval and pre-ranking stage are completed, the pre-modeling result in the cache is fetched and sent to the mid-modeling module along with pre rank result, mid-modeling module predicts all targets score. {\bfseries3) Post-Modeling.} a post-modeling stage is added, for different business scenarios, the sorting strategies could be quite different. The post-modeling stage performs personalized post-processing on the predicted items to meet business requirements, again, the pre-modeling result cached in redis is fetched as input parameters for post-modeling. The key used for storing pre-modeling results could be user id or request session id; the cached data life-cycle is configurable according to recommended accuracy and system cost.

\subsubsection{Design of Flexible Ranking Model.}
\begin{figure}[htbp]
\includegraphics[width=\textwidth]{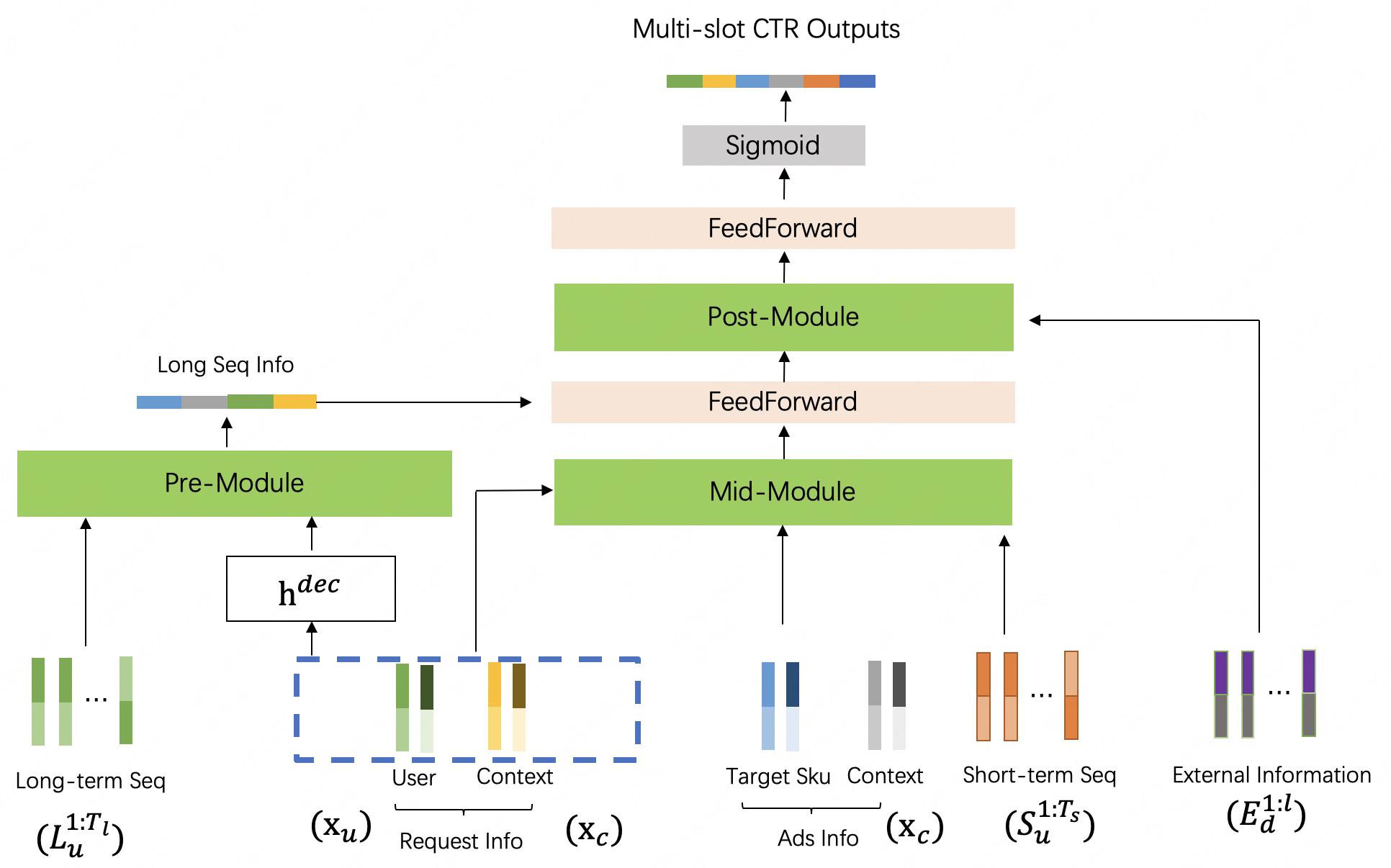}
\caption{Network Architecture of the CTR Model} \label{fig:model}
\end{figure}

Unlike the previous modeling, which reduces computing power cost by restricting model architecture and thus causes loss of model performance, PCDF allows applying arbitrary complex architecture of deep models to ensure the best model performance.   For example, in our real system, we take transformer based models \cite{vaswani2017attention} as our user behavior model architecture in pre-stage. Figure\ref{fig:model} illustrates the results from pre-stage  sent to  pos-model in  fully connected layer with the concatenation of other features  as inputs. Other model for user behavior and model for  other features independent with Ads,  are also applicable, which we leave readers for further trying.
in post-stage, the output from middle model is fused with the external items and compute the final score for the candicate items.

Multi-thread is also used for better concurrency performance, each thread accepts several user requests and business strategies will be applied to each request. Finally, a cache is used to reduce feature search latency and network communication costs.

\subsection{Deployment}
In this section, we give hands-on practice of the PCDF framework on online deployment in the industry. 

As mentioned above in Section  \ref{subsction:framework}, the model is split into three branches when online serving: long-term behavior sequence module deployed in pre-modeling module, the rest candidate item-dependent models deployed in mid-modeling module and modeling external information in pos-modeling module.

Although three sub-models are responsible for different tasks and are called sequentially, we export one dynamic computation graph and deploy the whole graph on the same server. The Prediction Server can choose the PCDF or CTR branch output corresponding to the request. Specifically, the Prediction Server can know the rank stage from the requests sent by the interface  Server. This deployment method naturally supports online learning since there is actually only one serving computation graph. Furthermore, deployment on the same machine contributes to consistency and enables easy management of all model versions, e.g., rollback or model structure updating. 

\begin{figure}[htbp]
\centering\includegraphics[width=0.6\linewidth]{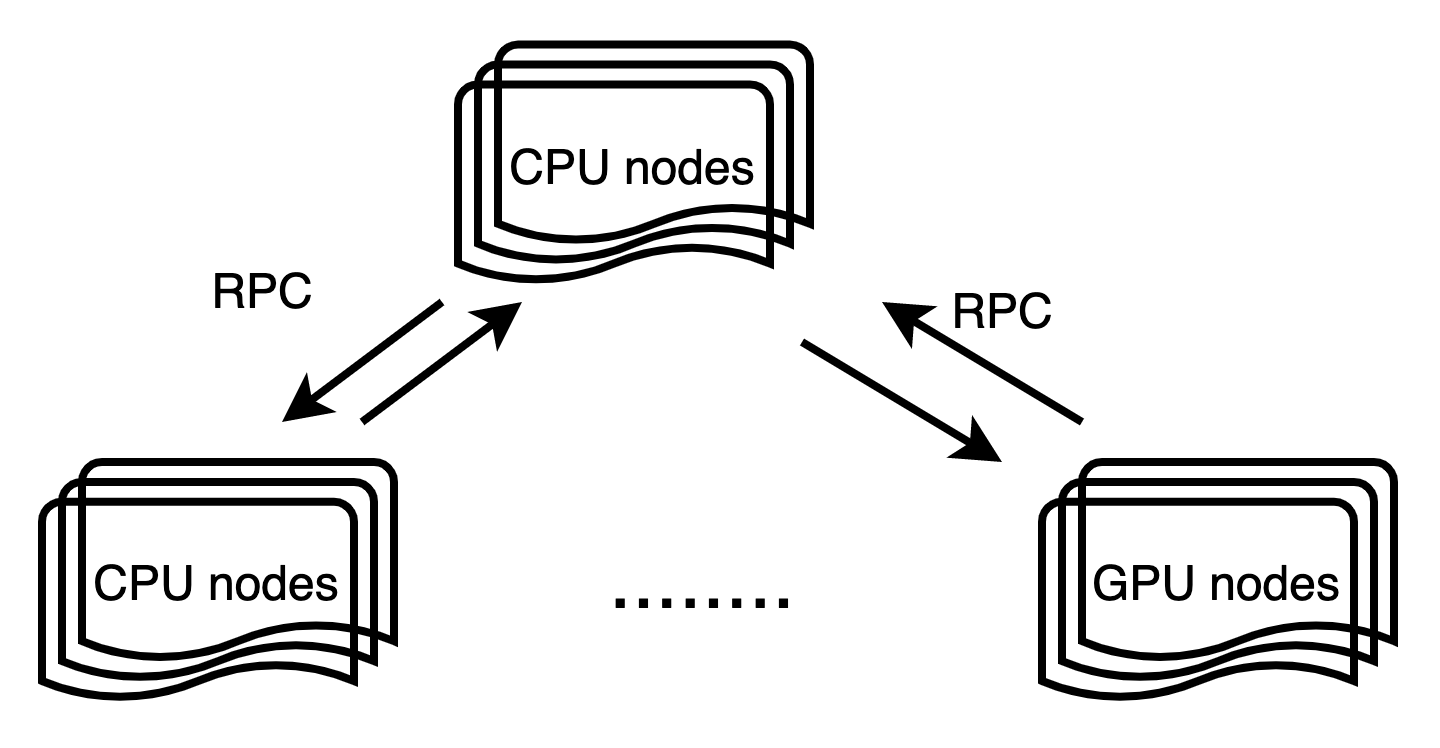}
\caption{CPU and GPU Isolation.} \label{fig:nodes}
\end{figure}

{\bfseries Isolation of CPU and GPU computing resources.} As mentioned in \ref{sec:consider}, it's hard to balance CPU resources and CPU resources in large-scale recommendation systems. To reduce the difficulty of deployment, we treated recommendation processes as computationally intensive and io-intensive, shown in figure \ref{fig:nodes}. we designed a distributed computing architecture that isolates CPU resources from GPU resources and split them to different nodes. RPC(Remote Procedure Call) is applied to exchange data between GPU nodes and CPU nodes.  Using appropriate computing graph splitting strategy, hash operation and request unpacking are handled by CPU nodes while model inference is performed on GPU nodes. As the model changes, CPU and GPU computing resources are adjusted independently. This distributed recommendation system architecture can greatly alleviate waste of computing resources in large industrial scenarios. Our hands-on practice shows that resource utilization increases from 35\% to 65\%. It is expected that as the model calculation distribution continues to change, the resource utilization rate will be further improved.

{\bfseries Other Optimization Trick.} Apart from pipeline parallelism mentioned in section \ref{introduction}, which is good at accelerating target-independent modeling, we also apply various data parallelism optimizations that may fit most recommendation processes. In our system, the target item's score ranking process is independent of each other. That means score computing could be processed in parallel with each other. At the front-end use request level, each request will be split into several inference sub-requests; each sub-request handles part of targets, after all sub-request processes are finished, results will be merged and ranked by score. The trade-off will be made when split user request since RPC is used in our recommendation system, too many RPC network communications means sub-requests have more chance get failed.

\section{Experiments}
We carefully compare the prediction latency of the baseline and our proposed framework at varying lengths of behavioral sequences. Furthermore, we compare the proposed search framework with several state-of-the-art works in modeling long-term behaviors on an industrial dataset. Finally, we conduct an A/B online test to verify the performance of PCDF.

\subsection{The Impact of The Sequence Length}
For validating the newly designed framework’s effectiveness, we conduct the prediction latency experiments carefully in our online advertising system.

\subsubsection{Experimental Settings}
For comparison of the impact of behavior lengths between the baseline and our new framework, we adopt precisely the same model in Section \ref{subsction:model} and denote \textit{Baseline} and \textit{PCDF} as the deployment method mentioned in \ref{sec:consider} respectively. 
Precisely, the \textbf{\textit{Baseline}} deploys the whole CTR model in Deep Rank module while \textbf{\textit{PCDF}} deploys long-term modeling in pre-computing module and the rest in Ranking module. 
Both the baseline and PCDF deployed the same hardware environment. retrieval server runs on a machine with 1 Intel Xeon CPU E5-2683  and 8GB RAM, and deep rank module Server runs on a machine with  1 Intel Xeon Gold 6267C CPU and 128GB RAM; The connection bandwidth between each service is 10Gbps.


\begin{figure}[htbp]	
	\subfigure[deep rank latency] 
	{
            \label{fig:subfigure5a}
		\begin{minipage}{0.5\textwidth}
			\centering          
			\includegraphics[scale=0.32]{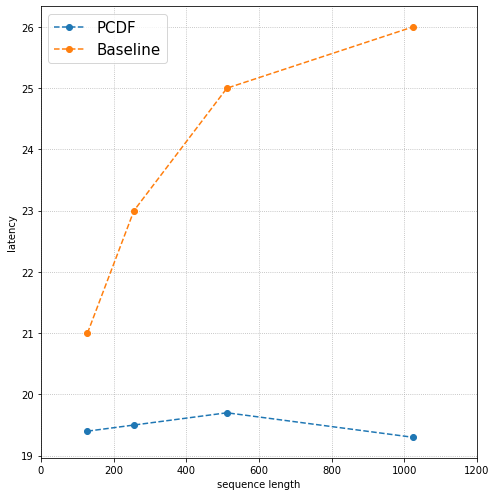}   
		\end{minipage}
	}
	\subfigure[Total latency] 
	{
            \label{fig:subfigure5b}
		\begin{minipage}{0.5\textwidth}
			\centering      
			\includegraphics[scale=0.37]{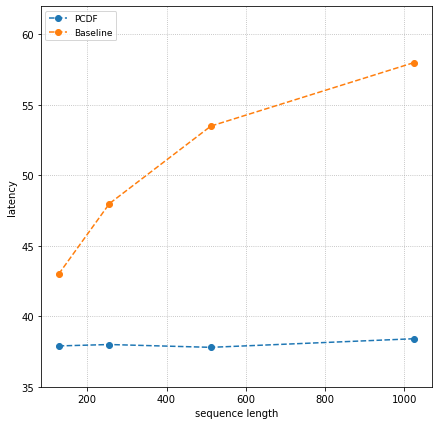}   
		\end{minipage}
	}
	\caption{Illustration of system latency, {\bfseries(a)} shows that baseline CTR prediction latency increase as behavior sequences increases, while PCDF remains stable, {\bfseries(b)} shows total latency of deep ranking stage, the latency trend is consistent with {\bfseries(a)}} 
\end{figure}


\subsubsection{Prediction Latency Results}
\textit{PCDF} achieve better performance on latency compared with \textit{Baseline}, demonstrating its high efficientness.
Figure \ref{fig:subfigure5a} gives the overall latency of the \textit{PCDF} and \textit{Baseline} in the ranking stage under different behavior lengths. With an increase in the length of behavior sequences, the overall latency of \textit{Baseline} shows an upward trend, where the latency increases by 15ms when the length goes up from 128 to 1024.
In contrast, the latency in the ranking stage of \textit{PCDF} remains stable at about 38ms, even though the length reaches 1024. It is noted that the predictor latency of \textit{Baseline} is about 58ms, which is not acceptable considering our strict online latency.

The results show that the latency of the baseline in the ranking stage keeps increasing with increasing sequence length, while the latency of PCDF is stable. It is noticed that about 60ms prediction latency in the original framework when length at 1024, shown in Figure \ref{fig:subfigure5b}, is unable to deploy under the constraint of extremely low system latency. Therefore, PCDF provides a solution and enables our complex modules like long-term user behavior modeling deployment. 

\subsection{Experiments  on Industrial Dataset}
\subsubsection{Competitive Models.} 
 {\bfseries SIM} \cite{ref_sim} is a CTR prediction model, which proposes a search unit to extract user interests from long-term user behavior sequences in a two-stage manner. SIM(hard) is the SIM that searches top-k behavior items by category id in the first stage. We follow previous work to compare SIM(hard) as the performance is almost the same as they deploy SIM(hard) online.  
{\bfseries ETA}\cite{chen2021end} applies LSH to encode target items and user behavior sequences into binary codes and then calculates the Hamming distance of the items to select top-k similar items for subsequent target attention  in an end-to-end manner.
\subsubsection{Experimental Settings}
For all the baselines and PCDF, we use the same features as input and adopt the same model structure except for the long-term user behavior modeling module.  All models use the same length of long-term user behaviors.
\subsubsection{Industrial Dataset.} We select consecutive 15-day samples for training and the next two days for  evaluation and testing, the number of training examples is about 10 billion. The recent 50 behaviors are selected as short-term sequences, and the recent 1,024 behaviors are selected as long-term sequences. 

\subsubsection{Result on offline experiment}
The evaluation results are shown in Table \ref{tab:exp_indus}. For industrial dataset, we report the result of  baselines and  PCDF. The best performance is highlighted in bold.
Results show that PCDF consistently outperforms all baselines on the industrial dataset. Specifically, it achieves improvements over the strongest baselines in terms of AUC  by 2.51\%, 1.60\%. This validates the effectiveness of our  module for modeling long-term historical behaviors for CTR prediction task. Note that SIM(hard) performs worst in the baseline models due to the loss of information caused by user behavior retrieval according to category.

\begin{table}
  \centering
  \caption{ Performance comparison of PCDF to other  methods on modeling long-term  behavior sequence}
  \label{tab:exp_indus}
  \begin{tabular}{lc}
    \toprule
     Model\quad \quad \quad \quad \quad \quad  \quad \quad &AUC\\
    \midrule
    SIM(hard)\quad \quad \quad \quad \quad \quad  \quad \quad  &0.7290\\
    ETA\quad \quad \quad \quad \quad \quad  \quad \quad &0.7355\\
    PCDF\quad \quad \quad \quad \quad \quad  \quad \quad &{\bfseries 0.7473}\\
  \bottomrule
\end{tabular}
\end{table}

\subsection{Online A/B Test}
The PCDF  is deployed in our real display advertising system with a long-term behavior sequence module with a length of 1024 and modeling the external information.

A strict online A/B test is conducted. Table \ref{tab:ab}  shows online experimental results using the proposed PCDF framework while the original framework without long-term behavior module and post-module as a benchmark. Compared to the benchmark, the PCDF achieves  5\% CTR and 5.1\% RPM (Revenue Per Mille) gain. 
Note that 5.1\% improvement on RPM is nontrivial improvement given that all other components of our base model have already been highly optimized and this leads to additional millions of revenues per day.

Finally, we analyze the efficiency of the proposed PCDF in online serving. The comparison of latency is also examined at the peak of queries per second (QPS), as shown in Table \ref{tab:ab}.
Adding long-term behavior modeling and post-modelling almost brings no extra inference time compared with the base model, proving the stability and effectiveness of our proposed framework.

We tested the performance of the pre-model and the post-model online respectively. Among them, the effect of the long-term behavior modeling  in  the pre-module achieves  3\% CTR and 3.1\% RPM gain and the  post-module modeling of organic search  information achieves  2\% CTR and 2\% RPM gain ,respectively. All of them have little to no time-consuming addition to the overall  ranking stage.

\begin{table}
  \centering
  \caption{ Online effectiveness and efficiency of the proposed framework on A/B test}
  \label{tab:ab}
  \begin{tabular}{cccc}
    \toprule
     &CTR&RPM&latency in ranking stage\\
    \midrule
    PCDF Framework & +5.00\% & +5.10\%&+0.4 ms\\
  \bottomrule
\end{tabular}
\end{table}

\section{Conclusions}
In this paper, we propose an efficient framework PCDF for deploying complex models and utilizing personalized externalities in CTR prediction for practical e-commerce sponsored search systems. The framework is designed for online learning through joint training and online distributed deployment, without increasing extra online inference time for complex computation modules. We comprehensively present the rationale behind our deployment, offer practical experience and conduct various experiments to demonstrate the superiority of our proposed PCDF. To the best of our knowledge, our work represents the first systematic solution to address the unacceptable efficiency of computationally intensive modules in CTR prediction, thus creating a new research direction.
\bibliographystyle{splncs04}
\bibliography{mybibliography}



\end{sloppypar}
\end{document}